\def\year{2021}\relax
\documentclass[letterpaper]{article} 
\usepackage{aaai21}  
\usepackage{times}  
\usepackage{helvet} 
\usepackage{courier}  
\usepackage[hyphens]{url}  
\usepackage{graphicx} 
\usepackage{graphicx}  
\usepackage{natbib}  
\usepackage{caption} 
\usepackage{subfigure}
\usepackage{booktabs}
\usepackage{multirow}
\usepackage{xcolor}
\usepackage[ruled]{algorithm2e}
\usepackage{array}
\usepackage{amsthm}
\usepackage{amsmath}
\usepackage{amssymb}
\usepackage{enumerate}
\usepackage[switch,columnwise]{lineno}
\theoremstyle{definition}
\newtheorem{definition}{Definition}

\newtheorem{prop}{Proposition}

\title{Efficient Optimal Selection for Composited Advertising Creatives\\ with Tree Structure}
\author{
  Jin Chen\thanks{This work was done when the authors Jin Chen and Gangwei Jiang were at Alibaba Group for intern.}$^1$, Tiezheng Ge$^2$, Gangwei Jiang$^3$, Zhiqiang Zhang$^2$, Defu Lian\thanks{Corresponding author}$^{3,4}$, Kai Zheng$^1$\\}
\title{Efficient Optimal Selection for Composited Advertising Creatives\\ with Tree Structure}
\author{
  Jin Chen\thanks{This work was done when the authors Jin Chen and Gangwei Jiang were at Alibaba Group for intern.}$^1$, Tiezheng Ge$^2$, Gangwei Jiang\footnotemark[1]$^{3}$, Zhiqiang Zhang$^2$, Defu Lian\thanks{Corresponding author}$^{3}$, Kai Zheng$^1$\\}
\affiliations{
  $^1$University of Electronic Science and Technology of China\\
  $^2$Alibaba Group\\
  $^3$University of Science and Technology of China\\
  chenjin@std.uestc.edu.cn, tiezheng.gtz@alibaba-inc.com, gwjiang@mail.ustc.edu.cn, \\
  zhang.zhiqiang@alibaba-inc.com, liandefu@ustc.edu.cn, zhengkai@uestc.edu.cn\\
}

\begin{document}

\maketitle

\begin{abstract}
Ad creatives are one of the prominent mediums for online e-commerce advertisements. Ad creatives with enjoyable visual appearance may increase the click-through rate (CTR) of products. Ad creatives are typically handcrafted by advertisers and then delivered to the advertising platforms for advertisement. In recent years, advertising platforms are capable of instantly compositing ad creatives with arbitrarily designated elements of each ingredient, so advertisers are only required to provide basic materials. While facilitating the advertisers, a great number of potential ad creatives can be composited, making it difficult to accurately estimate CTR for them given limited real-time feedback. To this end, we propose an \textbf{A}daptive and \textbf{E}fficient ad creative \textbf{S}election (AES) framework based on a tree structure. The tree structure on compositing ingredients enables dynamic programming for efficient ad creative selection on the basis of CTR. Due to limited feedback, the CTR estimator is usually of high variance. Exploration techniques based on Thompson sampling are widely used for reducing variances of the CTR estimator, alleviating feedback sparsity. Based on the tree structure, Thompson sampling is adapted with dynamic programming, leading to efficient exploration for potential ad creatives with the largest CTR. We finally evaluate the proposed algorithm on the synthetic dataset and the real-world dataset. The results show that our approach can outperform competing baselines in terms of convergence rate and overall CTR.
\end{abstract}

\section{Introduction} 
\begin{figure}[t]
    \centering
    \includegraphics[width=0.93\columnwidth]{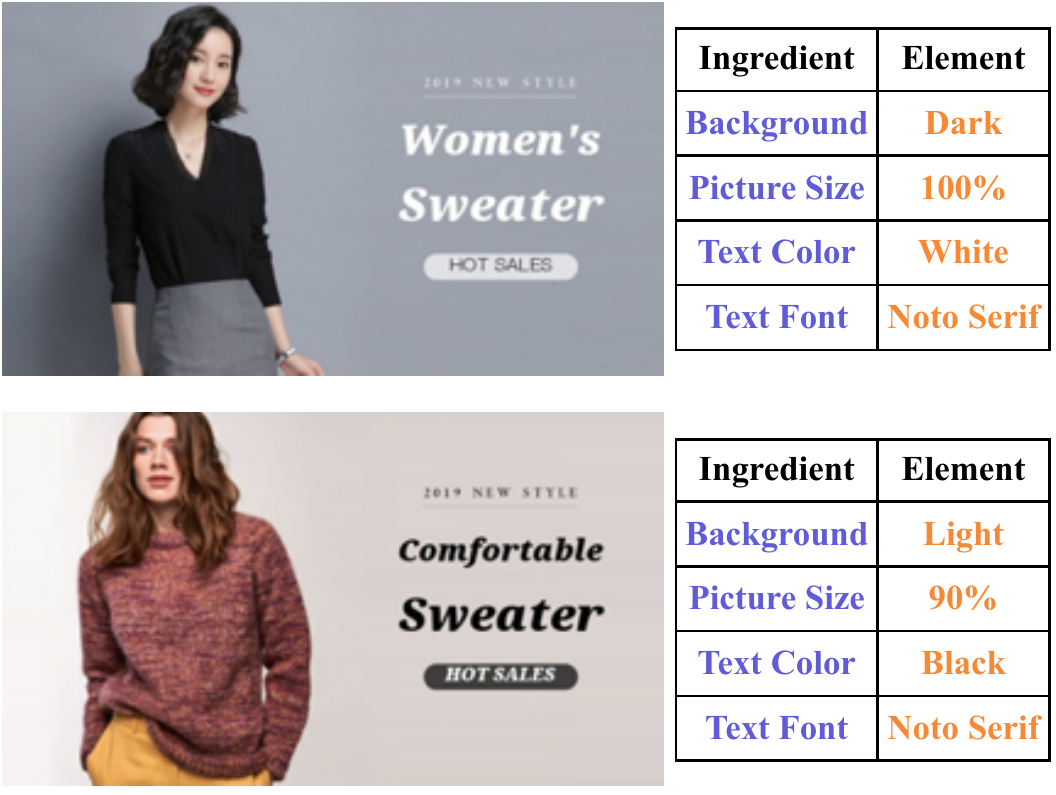}
    \caption{Ad creatives}
    \label{banners}
  \end{figure}

  With the rapid development of the Internet and mobile communication, the Internet has become the most important platform for advertisements, creating great commercial value for many high-tech companies via intelligent advertisements. Ad creatives, as shown in Fig.\ref{banners}, are one of the prominent mediums for online e-commerce advertisements. Ad creatives with enjoyable visual appearance may lead to a higher likelihood of click~\cite{mo2015image,teo2016adaptive}, and thus increase the click-through rate (CTR) of products. Several works~\cite{azimi2012impact,cheng2012multimedia} have shown that multimedia features of ad creatives, such as color histogram, color harmony and texts, are significantly correlated with the likelihood of user click. The increase of CTR may lead to an increase of revenue for e-commerce companies like Taobao and Amazon, so it is worth investigating how to improve visual appearance of ad creatives.

  Traditionally, ad creatives are handcrafted by advertisers and then delivered to advertising platforms for advertisement. Due to the small number of ad creatives, their optimal selection can be easily achieved by multi-arm bandit and variants in the advertising platform. However, advisers have to employ professional designers to design appealing ad creatives, incurring an additional economic cost. To help advertisers reduce the extra cost, some companies like Alibaba and Google establish intelligent platforms~\cite{hua2018challenges,kulkarni2018postview} for instantly compositing ad creatives with arbitrarily designated elements of each ingredient. The compositing ingredients usually include templates, text color, text font and picture size. The advertisers are only required to provide basic materials, such as product pictures and copywritings.

  However, the optimal selection for composited ad creatives is challenging. The reason lies in the following folds. First, multiple ingredients are composited combinatorially, leading to an exponential explosion in the number of ad creatives. For example, given 5 templates, 10 text fonts, 20 text colors and 20 picture sizes, 20,000 candidate ad creatives can be composited for a product. Moreover, the ingredients may change over time, so that ad creative spaces of each product are also varied. For example, the template often changes with advertising spaces, seasons and festivals. Second, due to the limited advertising budget of advertisers, the number of impressions for each product is usually small. According to our statistics in an advertisement platform of an E-commerce company, the number of impressions for each product within one day is less than 10 on average. When an impression of each item apportioned to its ad creatives, each ad creative is rarely impressed. Two questions are raised by these two challenges. First, given a large number of potential ad creatives, it is time-consuming to search the ad creative with the largest CTR. The necessity of online serving motivates us to develop efficient solutions. Second, given extremely sparse feedback, the estimated CTR for each ad creative is not accurate and of large variance, so that it may be imprecise to search the best ad creatives. 

  To this end, motivated by the sequential designing process of professional designers, we propose an \textbf{A}daptive and \textbf{E}fficient ad creative \textbf{S}election (AES) framework based on a tree structure over compositing ingredients. Upon the ingredient tree, dynamic programming is enabled for efficient selection of the best ad creative among all candidate creatives, while the accuracy of the CTR estimator is almost not affected by the tree structure assumption according to empirical study. This is extremely important when the CTR estimator is frequently updated. To reduce the variance of the CTR estimator incurred by sparse feedback, exploration techniques~\cite{montague1999reinforcement} based on Thompson sampling in multi-arm bandit are widely used for reducing the variance of the CTR estimator in online advertising~\cite{graepel2010web,chapelle2011empirical,agrawal2013thompson}. This is because a balance between exploration and exploitation in multi-arm bandit is stricken to minimize the opportunity cost of making sub-optimal decisions before knowing the best arm. Based on the ingredient tree, Thompson sampling is adapted with dynamic programming, leading to efficient exploration for the best ad creatives. Moreover, the ingredient tree structure makes it easy to incorporate visual constraints between elements of two ingredients. An example of visual constraints is that the light text color is not suitable for the light background color. 
    
  The contributions made in this paper are summarized as:
  \begin{itemize}
    \item To the best of our knowledge, this is the first attempt to investigate the optimal selection of composited ads-creatives in the scenario of E-commerce advertisements, and identify its unique challenges and research questions.
    \item We propose an \textbf{A}daptive and \textbf{E}fficient ad creative \textbf{S}election (AES) framework based on the ingredient tree, such that optimal selection is achieved by dynamic programming. We also adapt Thompson sampling with dynamic programming for efficient exploration.
    \item  We evaluate the proposed algorithm on the synthetic dataset and real-world dataset. The empirical results show the superiority of the proposed algorithm to the competing baselines in terms of convergence rate and overall CTR.
  \end{itemize}

  \section{Related Work}
  Regular CTR prediction on ad creatives~\cite{chen2016deep,liu2020category} has attracted a lot of attention for increasing CTR. However, optimal ads selection with sparse feedback is seldom studied.
  
  A similar task is combinatorial bandit~\cite{cesa2012combinatorial} focusing on selecting a set of arms at once. These works aim to select subsets of arms that fit in top-k problems~\cite{rejwan2020top}, such as selecting web pages for displaying~\cite{chen2013combinatorial}. However, we aim to select the best creative instead of subset.
  
  A large number of bandit algorithms depending on structured arms have been implemented for online advertising and personalized recommendation~\cite{chapelle2011empirical,li2010exploitation}. Existing works propose tree structures for fast search in the form of hierarchies where the arm is a leaf vertex of the constructed tree~\cite{bubeck2010open}. Dependencies among arms are formulated in the form of hierarchies~\cite{wang2018online} and applied for online recommendation. CoFinUCB~\cite{yue2012hierarchical} encodes prior knowledge for quick exploration depending on coarse-to-fine feature hierarchy. However, the interactions between candidate elements in our problem can not be embedded as hierarchy structures.

  \section{Optimal Selection of Ad creatives with Classical Methods}

  We have mentioned that we only consider ad creatives composited with ingredients rather than designed by designers, where the ingredient is defined as follows:
  
  \begin{definition}[Ingredient]
    Ingredients are basic components used for compositing ad creatives, including template, text color, text size, background color, picture size and so on.
  \end{definition}
  Each ingredient is actually a collection of \emph{elements}, where real-valued ingredients are also discretized for convenience. For designated ingredients and materials such as product images and copywritings, \emph{the composition of them into ad creatives is achieved by first selecting elements in each ingredient and then synthesizing ad creatives with the materials and chosen elements.} Only the first problem is the focus of this paper, and the second one has been studied in computer vision~\cite{li2019layoutgan}. Some examples of ad creatives and composting ingredients are shown in Fig.~\ref{banners}.  
  
  Ideally speaking, to achieve the optimal selection of ad creatives, we should enumerate all possible ad creatives for each product, and deliver them to the advertising platform to collect feedback for estimating the click-through rate (CTR). We then select the ad creative with the maximal CTR for future advertisements.

  \subsection{Multi-Arm Bandit}
   One classical method for optimal creative selection is multi-arm bandit by drawing an analogy between bandits and creatives. Multi-arm bandit methods have been widely used in many similar optimal selection tasks, such as the whole-page optimization~\cite{tang2013automatic,wang2016beyond}. These algorithms proceed in discrete time steps $t=1,2,...,T$. At time step $t$, an ad creative ${\mathbf{c}}_t$ is selected for advertisement, and then a reward $R_{{\mathbf{c}}_{t}}\in\{0,1\}$ is returned in each impression. An impression occurs when an ad creative is exposed to a user, and $R_{{\mathbf{c}}_{t}}=1$ indicates the user clicks the advertising creatives. The objective of the multi-arm bandit is to minimize the cumulative regrets within $T$ steps:
   \begin{displaymath}
     \min \sum_t^{T} ( R_{\mathbf{\mathbf{c}}^{*}} - R_{\mathbf{\mathbf{c}}_{t}}) 
   \end{displaymath}
  where $\mathbf{\mathbf{c}}^* = \arg\max \mathbb{E}(R_{\mathbf{\mathbf{c}}})$ denotes the candidate creative with maximum expected reward. UCB~\cite{auer2002using} is an effective method for balancing the trade-off between exploration and exploitation in multi-arm bandit, by measuring the potential with an upper confidence bound of reward. However, since creatives are composited by elements of ingredients, it is necessary to leverage commonality between creatives for estimating reward. LinearUCB~\cite{chu2011contextual} can be used in this case, which models the expected reward of each creative as a linear function of context features $ \mathbf{x}_{\mathbf{\mathbf{c}}}$ (the selected elements of the creative):
  \begin{displaymath}
    \mathbb{E}(R_{{\mathbf{c}}}) = \mathbf{x}_{{\mathbf{c}}}^\top \theta
  \end{displaymath}
  where $\theta$ is a parameter to estimate. The parameter will be updated after each impression is received. Similar to UCB, LinearUCB selects the creative with the maximal value of $\mathbf{x}_{\mathbf{\mathbf{c}}}^T {\theta} + \alpha \sqrt{\mathbf{x}_{\mathbf{\mathbf{c}}}^T \mathbf{A}^{-1} \mathbf{x}_{\mathbf{\mathbf{c}}}}$, where $\alpha$ is the coefficient for exploration and $\mathbf{A}=\sum_t \mathbf{x}_{\mathbf{\mathbf{c}}_t}^T \mathbf{x}_{\mathbf{\mathbf{c}}_t} $.
  
  \subsection{Multivariate Bandit}
  
  Assume there are $N$ ingredients to be composited, and the $i$-th ingredient $I_i$ has $L_i$ elements. Then the number of candidate creatives is up to $L_1\times L_2 \times ... \times L_N$, leading to an exponential explosion. Therefore, it is inefficient to enumerate all potential creatives and apply multi-arm bandit for optimal selection. A more efficient method is multivariate bandit~\cite{hill2017efficient}, which models interactions between elements of two different ingredients explicitly by defining the expected reward for each creative as follows:
  \begin{equation}
  \mathbf{x}_{\mathbf{\mathbf{c}}}^T \theta = \theta^0 + \sum_{i=1}^D \theta_i^1(\mathbf{\mathbf{c}}) + \sum_{j=1}^D \sum_{k=j+1}^D \theta^2_{j,k}(\mathbf{\mathbf{c}})	
  \end{equation}
  where $\theta^0$ is a common bias, $\theta^1_i(\mathbf{\mathbf{c}})$ is a weight associated with the element in the $i$-th ingredient, and $\theta^2_{j,k}$ is a weight for the interaction between the elements of the $j$-th and $k$-th ingredients. Nevertheless, searching the ad creative with the maximal reward is very time-consuming, so that hill climbing is applied for the approximated optimal creative selection. Moreover, instead of upper confidence bound used in multi-arm bandit, Thompson sampling was used for efficient exploration.
  
  However, several issues in the multivariate bandit remain to be solved. First, the multivariate bandit considers any possible interactions between elements so that it can not incorporate visual constraints, which is important for creating appealing ad creatives. Second and more importantly, although hill climbing can reduce the time cost of searching, it is almost impossible to find out the globally best creatives. And it is also possible for hill climbing to return creatives which do not satisfy the visual constraints. These issues motivate us to design a more powerful algorithm for efficient optimal selection of ad creative in case of large creative space.

  \begin{figure}[t]
    \centering
    \includegraphics[width=\columnwidth]{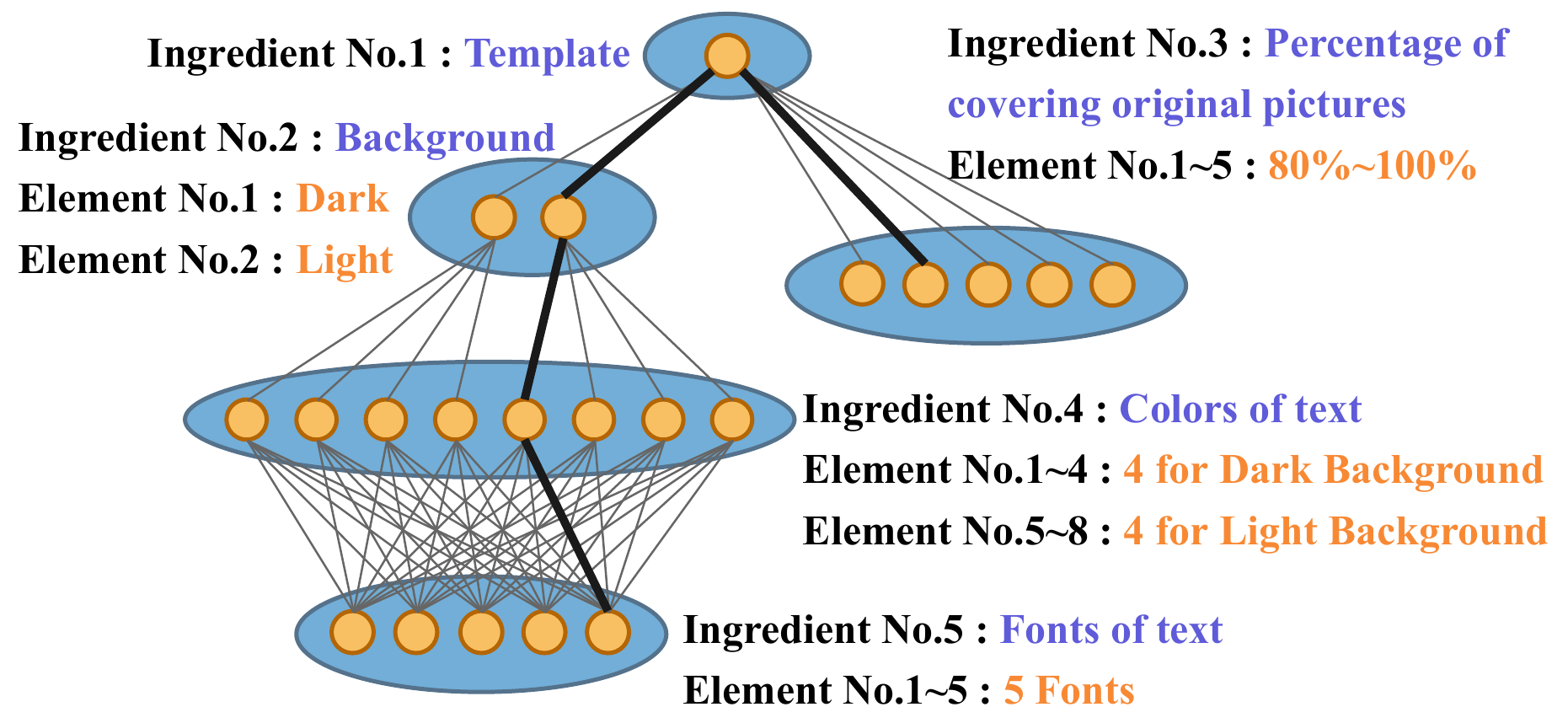}
    \caption{Ingredient Tree and element graph. Vertices in blue constitute the Ingredient Tree. Vertices in yellow constitute the element graph. An example feasible ad creative is shown as the sub graph with the connected bold edges.}
    \label{banner_structure}
  \end{figure}
  
  \section{Efficient Optimal Selection of Ad creatives with Ingredient Tree}

  In this section, we first introduce the ingredient tree and element graph, and then investigate how to perform an efficient optimal selection of composited ad creatives.
  \subsection{Preliminary}

  For the sake of efficient optimal selection among the large creative space, we introduce an ingredient tree, which is defined as follows:
  \begin{definition}[Ingredient Tree]
    Let $\mathcal{T}=\left( V^I, E^I \right)$ denote the Ingredient Tree where $V^I=\{I_1,\cdots, I_N\}$ is the set of $N$ ingredients and $E^I$ represents the relationships between ingredients.
  \end{definition}
  An example of the ingredient tree is shown in Fig.~\ref{banner_structure}. Actually, the ingredient tree is also motivated by imitating the decision making process of designing ad creatives from the perspective of professional designers, where ingredient is specified from top to bottom in the tree for compositing ad creative.
  
  To enable searching in the ingredient tree, we have to specify the element graph based on the ingredient tree, where each ingredient is substituted with its elements, as also shown in Fig.~\ref{banner_structure}.    
  
  \begin{figure}[t]
    \centering
    \subfigure[Good case]{
      \begin{minipage}{0.47\columnwidth}
        \includegraphics[width=1.55in]{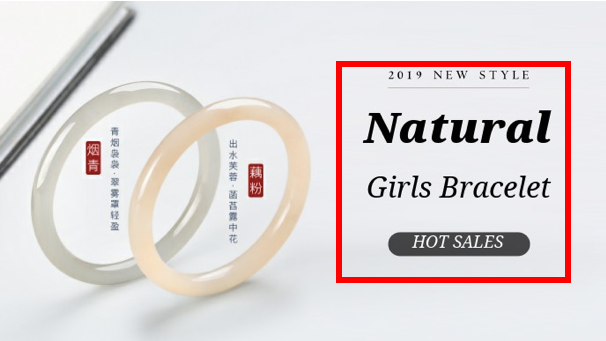}
      \end{minipage}
    }
    \subfigure[Bad case]{
      \begin{minipage}{0.47\columnwidth}
        \includegraphics[width=1.55in]{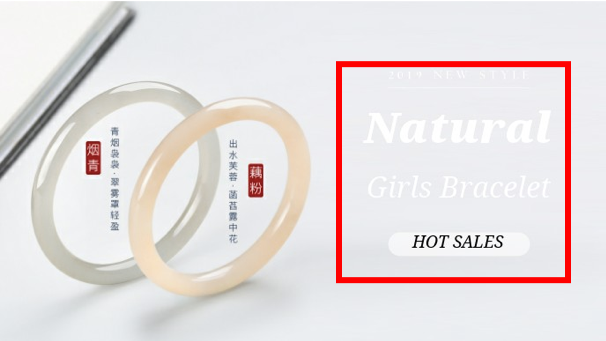}
      \end{minipage}
    }
    \caption{Candidate creatives for one example item. Best viewed in color. (a) Light background with the dark color of texts is a good case. (b) As shown in red box, copywritings are not easy to follow under the combination of the light background and the light color of texts.}
    \label{show_not_full}
  \end{figure}

  \begin{definition}[Element Graph]
    $\mathcal{G}^E=\left(V^E,E^E\right)$ is an element graph, where $V^E=I_1\cup \cdots \cup I_N$ is the set of elements of $N$ ingredients and $E^E$ is the set of edges, indicating the matching level between elements of two ingredients.
  \end{definition}
  Thanks to no connection between two elements in an ingredient, the element graph has a special property, which can be used for accelerating ad creative search.
  \begin{prop}
    The subgraph corresponding to any two connected ingredients in the ingredient tree is a bipartite graph.
  \end{prop}
  Actually, the whole element graph is also a bipartite graph. The number of nodes in the element graph equals to $|V^E|=\sum_{i=1}^N L_i$. Given the element graph, it is very convenient for us to incorporate visual constraints, such as the bad case shown in Fig.\ref{show_not_full}. These constraints are incorporated by removing connections between certain elements of both the \textit{picture} ingredient and the \textit{text color} ingredient, indicating the 0-level of matching between them. 
  
  \subsection{Overview}
  Given the ingredient tree and element graph, we propose an \textbf{A}daptive and \textbf{E}fficient ad creative \textbf{S}election (AES) framework. The proposed framework consists of the following three phases:
  
  \begin{enumerate}[1)]
    \item \textbf{CTR Estimation:} The CTR estimator is built by considering weights of edges and vertices in the element graph as parameters, based on which the ad creative with the largest CTR will be selected in the next phases.
    \item \textbf{Optimal Creative Selection:} Dynamic programming is proposed on the element graph for efficiently searching the best ad creative, aiming to dramatically reduce the time cost of searching.
    \item \textbf{Efficient Exploration:} The classical exploration method -- Thompson sampling is adapted with dynamic programming to improve the efficiency of exploration and to address feedback sparsity.
  \end{enumerate}
  
  \subsection{CTR Estimation}
  
  Based on the element graph, by selecting an element in each ingredient, we can construct a feasible candidate creative. It can be more formally defined as follows:
  \begin{definition}[Candidate Creative]
    A feasible candidate creative $\mathbf{c}$ corresponds to a connected subgraph $\mathcal{G}^E_{\mathbf{c}}=(V^E_{\mathbf{c}},E^E_{\mathbf{c}})$ of the element graph $\mathcal{G}^E$, being isomorphic to the ingredient tree $\mathcal{T}$.
  \end{definition}
  The example of a feasible creative corresponds to the black bold line in Fig.~\ref{banner_structure}. As aforementioned, weights of edges and vertices in the element graph are considered as parameters, so similar to contextual bandit, the expected reward of a candidate creative $\mathbf{c}$ is formulated into a linear function with respect to vertices and edges of its subgraph:
  \begin{equation}
      \mathbb{E}(R_{{\mathbf{c}}}) = b + \sum_{j\in V^E_{{\mathbf{c}}}} w_j + \sum_{(i,j) \in E^E_{{\mathbf{c}}}} w_{i,j}
      \label{ctr_linear_assum}
  \end{equation}
  where $b$ is a bias term, $w_j$ is a weight of the selected element $j$ in the creative $\mathbf{c}$, and $w_{i,j}$ is the weight of the edge between selected element $i$ and selected element $j$ in the creative $\mathbf{\mathbf{c}}$. $V^E_{\mathbf{\mathbf{c}}}$ and $E^E_{\mathbf{\mathbf{c}}}$ is the set of vertices(elements), and the set of edges between elements of connected ingredient in the ad creative ${\mathbf{c}}$. The number of parameters in the linear function is $|V^E|+|E^E|$ in total. Since these parameters are estimated with the click-through logs of ad creatives, the reward of each creative is also called click-through rate (CTR). As mentioned, the CTR estimator will be used for optimal selection of ad creatives.
  
  \subsection{Optimal Creative Selection}
  
  \begin{algorithm}[t]
    \caption{\small{Dynamic Programming for Ads Selection}}
    \label{explore_after_dp}
    \LinesNumbered
    \KwIn{Ingredient Tree $\mathcal{T}$, Element graph $\mathcal{G}^E$, vertex weights $\mathbf{w}$, edge weights $\mathbf{v}$}
    \KwOut{Optimal solution}
  
    \SetKwFunction{DS}{DynamicSearch}
    \SetKwProg{Fn}{Function}{:}{}
    
    \Fn{\DS{$I_i$}}{
      \textit{$ \backslash \backslash I_i$ is an ingredient}\;
      Initialize $\mathbf{d}_{I_i}\leftarrow \mathbf{0}_{L_i}$\;
      \For{element $j \in I_i$}{
          $\mathbf{d}_{I_i}[j] = w_j$\;
        }
      \If{$I_i$ is the Leaf ingredient}{
        \KwRet $\mathbf{d}_{I_i}$\;
      }
      \For{$I_m \in \{ Child[I_i]\}$}{
        $\mathbf{d}_{I_m}=$ \DS{$I_m$}\;
      }
      \For{element $j \in I_i$}{
        \For{$I_m \in \{ Child[I_i]\}$}{
          $\mathbf{d}_{I_i}[j] = \mathbf{d}_{I_i}[j] + \max_{t\in I_m}(v_{i,t}+\mathbf{d}_{I_m}[t])$
        }
      }
      \KwRet $\mathbf{d}_{I_i}$\;
    }  
    Run $\mathbf{d}_{\mathcal{R}}=$ \DS{$\mathcal{R}$}\;
    $\backslash \backslash \mathcal{R}$ is the root ingredient of $\mathcal{T}$\;
    Rerurn $\max_{j\in \mathcal{R}}\mathbf{d}_{\mathcal{R}}[j]$ as the optimal solution\;
    
  \end{algorithm}
  Before delving into the algorithm, we first define the optimal creative based on candidate creatives as follows:
  \begin{definition}[Optimal Creative]
    The optimal creative is a feasible candidate with the largest estimated CTR.
  \end{definition}
  Since edge weight indicates the matching/consistent level between elements of two ingredients, the optimal creative is considered as globally best configured with the ingredients. Searching the optimal creative by enumerating all ad creatives suffers from computational challenges, since the number of feasible creatives is still large even with the ingredient tree assumption. However, given the element graph defined over the ingredient tree, it is possible to develop dynamic programming for searching the optimal creatives even when the CTR estimator is frequently updated, so that search efficiency can be remarkably improved.

  As mentioned, in the element graph, $w_i$ represents the weight of vertex $i$ and $v_{i,j}$ denotes the weight of edge between two connected elements $i$ and $j$. Assume each ingredient $I_i$ is associated with a vector $\mathbf{d}_{I_i}$ of size $L_{i}$. Each entry $\mathbf{d}_{I_i}[j]$ in the vector $\mathbf{d}_{I_i}$ denotes the maximum value of the subtree rooted at the element vertex $j$. Then the optimal solution equals to $\max_{j}{\mathbf{d}_{\mathcal{R}}[j]}$ where $\mathcal{R}$ denotes the root ingredient in the ingredient tree. The state transition equation for dynamic programming is followed as:
 
  \begin{displaymath}
    \small
    \mathbf{d}_{I_i}[j] = \left\{
      \begin{aligned}
        & w_j , & if \,\, I_i \in \{\mathcal{L} \}\\
        & w_j + \sum_{I_m \in \{\mathcal{C}[I_i] \}} \max_{t \in I_m}{(v_{j,t} + \mathbf{d}_{I_m}[t])}, & else\\
      \end{aligned}
      \right.
  \end{displaymath}
  where $\mathcal{C}[I_i]$ denotes the set of \emph{child ingredients} of $I_i$. The element $t$ is an element in the child ingredient $I_m$. $\mathcal{L}$ denotes the leaf ingredients of the Ingredient Tree.
  The dynamic programming algorithm for optimum searching over the element graph is shown in Alg.~\ref{explore_after_dp}. The time complexity of this algorithm is $\mathcal{O}(|V^E|+|E^E|)$. $|V^E|$ is the number of vertices and $|E^E|$ is the number of edges in the element graph, which are usually significantly smaller than the number of candidate ad creatives.

\begin{algorithm}[t]
  \caption{AES (\textbf{A}daptive and \textbf{E}fficient ad creative \textbf{S}election framework)}
  \label{thompson_contextual}
  \LinesNumbered
  \KwIn{Element graph $\mathcal{G}^E$, prior variance $\sigma$, creative features $\mathbf{x} \in \mathbb{R}^{K},K=|V^E|+|E^E|$ }
  Initialize $\mathbf{B}\leftarrow\mathbf{I}_K$, $\mathbf{f}\leftarrow \mathbf{0}_M$, $\mathbf{\bar{w}} \leftarrow \mathbf{0}_K$\;
  \For{ t = 1,2,...,T}
  {
    Sample $\widetilde{\mathbf{w}}$ from $\mathcal{N}(\bar{\mathbf{w}},\sigma^2\mathbf{B}^{-1})$\;
    Find creative $\mathbf{\mathbf{c}}_t = \arg \max \mathbf{x}_{\mathbf{\mathbf{c}}}^T \widetilde{\mathbf{w}}$ with Alg.\ref{explore_after_dp}\;
    Recommend and receive feedback $R_{\mathbf{\mathbf{c}}_t}$\;
    Update $\mathbf{B} \leftarrow \mathbf{B} + \mathbf{x}_{\mathbf{\mathbf{c}}_t} \mathbf{x}_{\mathbf{\mathbf{c}}_t}^T$, $\mathbf{f} \leftarrow \mathbf{f} + R_{\mathbf{\mathbf{c}}_t} \mathbf{x}_{\mathbf{\mathbf{c}}_t}$, $\mathbf{\bar{w}} \leftarrow \mathbf{B}^{-1} \mathbf{f}$\;
  }
\end{algorithm}

\subsection{Exploration Methods}
Due to the limited advertising budget for advertisers and large numbers of candidates, each ad creative only receives a few impressions so that the CTR estimator is usually of high variance. Thus, we introduce efficient exploration methods to search potential good ad creatives for advertisement, with the aim of reducing the variance of estimated CTR.

Thompson sampling~\cite{thompson1933likelihood} is a popular bandit algorithm to balance exploitation and exploration. In our setting, this means the ad creative with not the currently highest expected reward (i.e., CTR) but potential maximal expected reward will be selected for advertisement. In practice, each candidate creative has a probability of being optimal based on previous impressions and Thompson Sampling selects a creative proportionally to the probability.

According to the CTR estimator we designed, by following Bayesian linear regression the posterior also follows Gaussian distribution if placing independent Gaussian prior on model parameters. Therefore, instead of directly sampling ad creative with the click probability, we sample from the posterior distribution of model parameters in the CTR estimator~\cite{agrawal2013thompson}. After the parameters are sampled, the ad creative with the highest CTR can be selected for advertisement. This indicates that dynamic programming can be applied in Thompson sampling so that the exploration of ad creatives can be very efficient. The \textbf{A}daptive and \textbf{E}fficient ad creative \textbf{S}election (AES) framework is shown in Alg.\ref{thompson_contextual}.


\section{Experiments}
In this section, we conduct experiments on both the synthetic dataset and the real-world dataset\footnote{https://github.com/alimama-creative/AES-Adaptive-and-Efficient-ad-creative-Selection} to evaluate our approach to answer the following questions: 
\textbf{Q1:} What is the efficiency and effectiveness of exploration when the tree structure is introduced?
\textbf{Q2:} What is the effect of the tree assumption over ingredients on the accuracy of the CTR estimator?
\textbf{Q3:} How the tree-based structure affects the efficiency of searching the best ad creative?

\subsubsection{Baseline algorithms}
We chose the following typical bandit algorithms for comparisons where parameters are tuned with best performances. 
  \begin{itemize}
    \label{Baseline}
    \item \textbf{Egreedy} Recommend a creative at random with probability $\epsilon$ and recommend the creative of best combination with probability $1-\epsilon$. We set $\epsilon=0.1$.
    \item \textbf{UCB} We recommend the creative with the maximum value of the upper bound~\cite{auer2002using} and $\lambda$ is set 0.03.
    \item \textbf{Ind-Egreedy} Independently perform Egreedy for each of $N$ ingredient.
    \item \textbf{LinearUCB} LinearUCB~\cite{chu2011contextual} assumes the linear interactions between the contextual features.
    \item \textbf{TEgreedy} Tree-Based Egreedy. We perform weighted linear regression depending on the tree-based estimator in Eq.\ref{ctr_linear_assum} and run Egreedy for exploration.
    \item \textbf{MVT} MVT~\cite{hill2017efficient} models possible interactions between elements. It utilizes hill climbing for optimal selection and thompson sampling for exploration.
  \end{itemize}

\subsection{Experiments on Synthetic Data}
\subsubsection{Simulated data}
\begin{figure}[t]
    \centering
    \includegraphics[width=0.95\columnwidth]{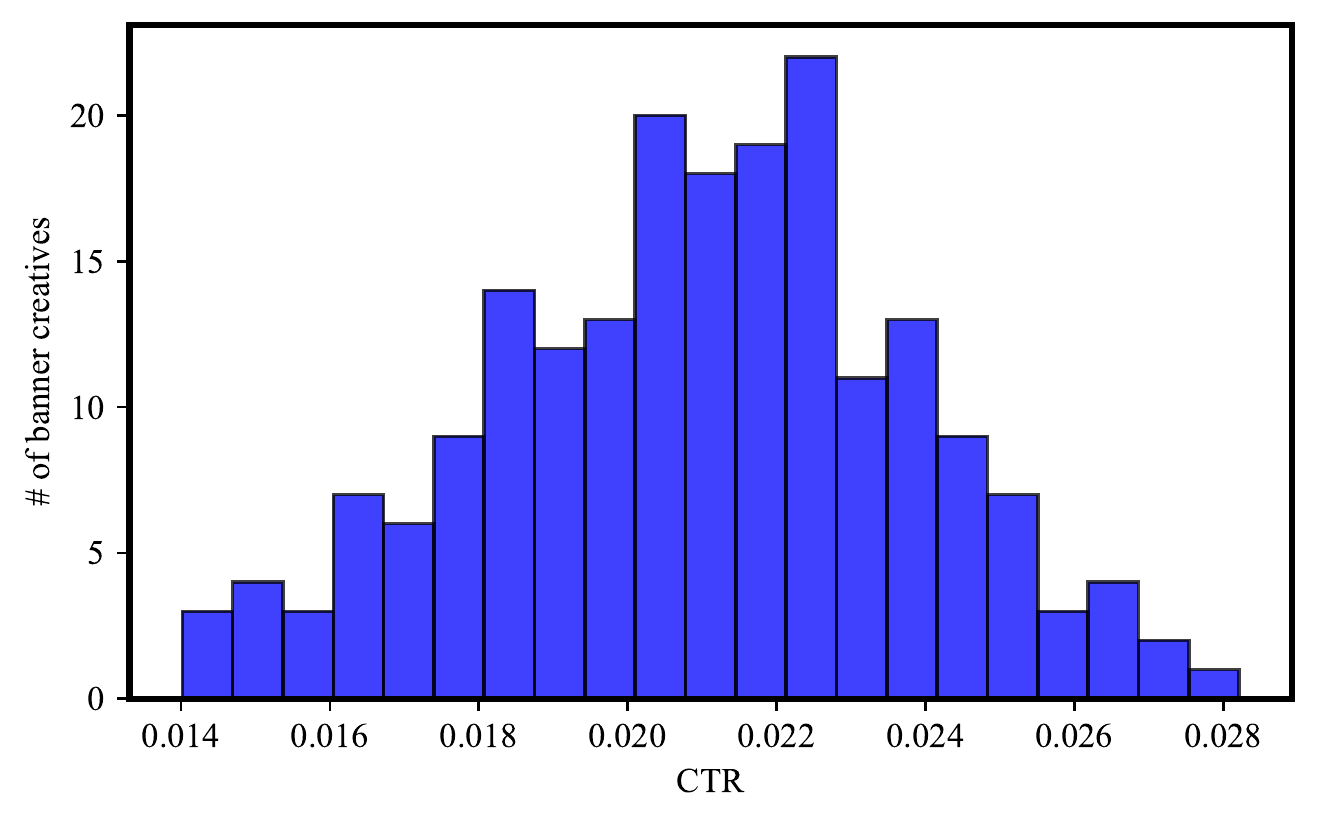}
    \caption{Histogram of simulated expected reward (CTR) for 200 ad creatives}
    \label{data_distribution}
\end{figure}

Following previous works, such as~\cite{hill2017efficient}, we produce synthetic data for experiment. The element graph we followed is shown in Fig.\ref{banner_structure}. There are five ingredients to be composited, including template, background of pictures, picture size, font and color of text, and 200 ad creatives are produced. We generate the expected reward for creatives following Eq.\ref{ctr_linear_assum}, where the weights of vertices and edges are sampled from a Gaussian distribution $\mathcal{N}(0,1)$. Fig.\ref{data_distribution} shows the distribution of expected reward in the simulation experiment. 
  
\subsubsection{Simulation settings}
We perform Bernoulli experiments to simulate the user feedback (click or not) for each impression, like previous work~\cite{fouche2019scaling}. When a creative $\mathbf{\mathbf{c}}$ is recommended, we perform a Bernoulli trial where the successful probability is the predefined expected reward of $\mathbf{\mathbf{c}}$. 

The models are trained every 1000 trials within one bacth to simulate the delayed feedbacks in the real-world system. To reduce the bias caused by randomness, we run each algorithm for 20 times. Our experiments are conducted in a Linux system with 256G memory and CPU E5-2682.

We take \textit{cumulative reward} (overall CTR) and \textit{regret}~\cite{hill2017efficient} as metrics for comparison. 

\subsubsection{Learning Effectiveness.} 
\begin{figure}[t]
    \centering
    \includegraphics[width=\columnwidth]{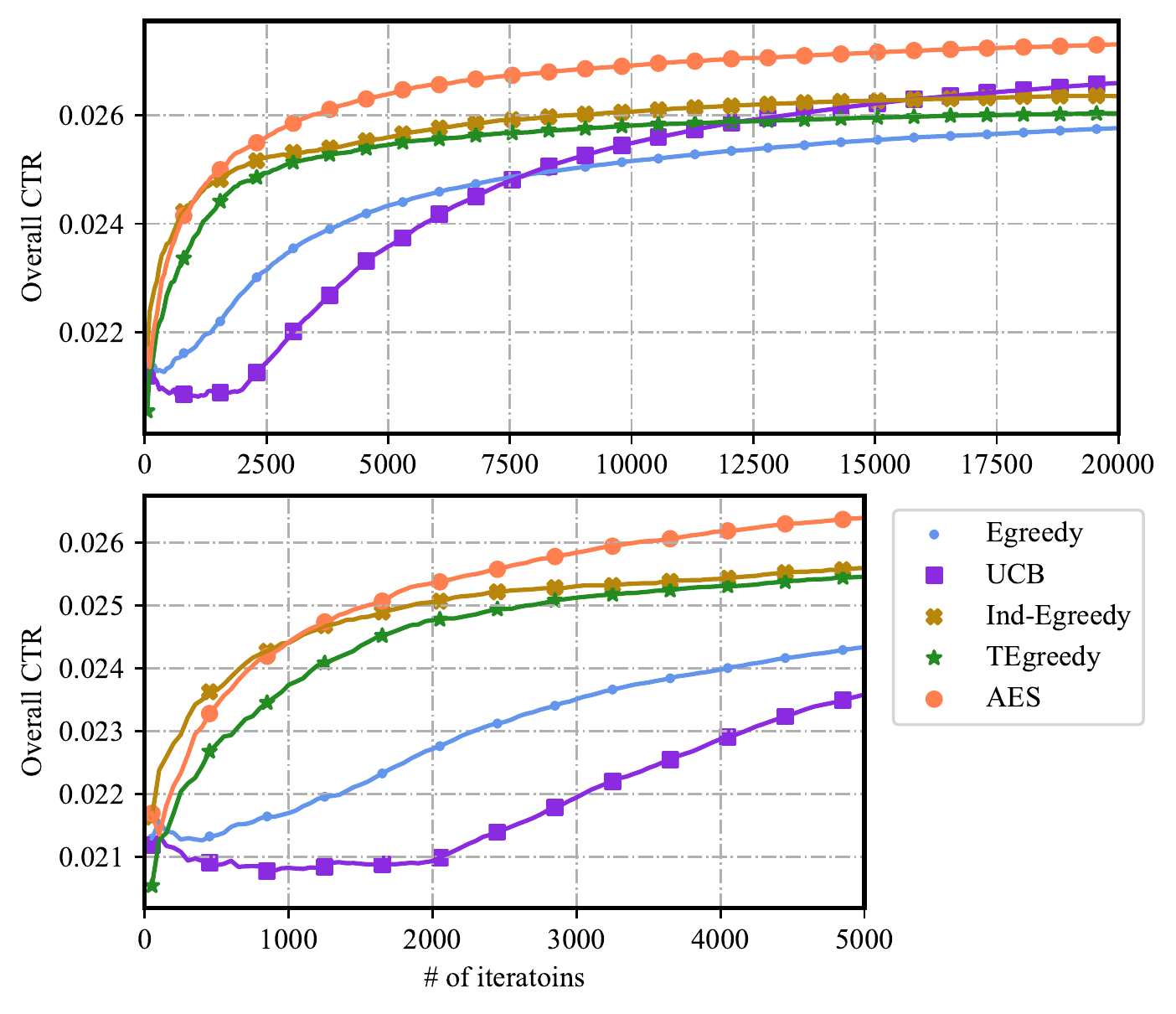}
    \caption{Result on simulated data. Performances within 500 batches are zoomed in (See in the figure below). Experiments are run for 20 repetitions.}
    \label{res_analyze}
\end{figure}

We perform 2,000 batches, i.e.2 millions impressions, for each algorithm to evaluate the capability of finding optimal composited creatives within limited impressions. Results are shown in Fig.\ref{res_analyze}.

\textit{Finding1}: Our approach introducing tree structures helps quickly find the best creative compared with context-free methods. Compared with Egreedy and UCB, our approach based on the ingredient tree, where connections of different elements are captured, shows continuously competitive performances, especially at the beginning 500 batches.

\textit{Finding2}: The tree-based structure benefits finding better creatives compared with Ind-Egreedy. Ind-Egreedy quickly increases the overall CTR but it finally converges to a lower point while AES has an increase of about 8.0\% relative to Ind-Egreedy after 200 batches. Ind-Egreedy method is a simple solution where each candidate ingredient is independently selected. Our method depending on the tree-based structure increases the overall CTR within a short time, which indicates the interactions between different ingredients also influence user engagements.

\textit{Finding3}: Various exploration methods can be flexiblly adapted into the tree structure and show good performances. TEreedy and AES are two adapted methods on the basis of the ingredient tree. The better performance of AES also shows that the exploration over the weights is more efficient than exploration over the whole creative space.

  \begin{figure*}[ht]
      \begin{minipage}{0.33\textwidth}
        \centering
        \captionsetup{margin=0.07in}
        \includegraphics[scale=0.47]{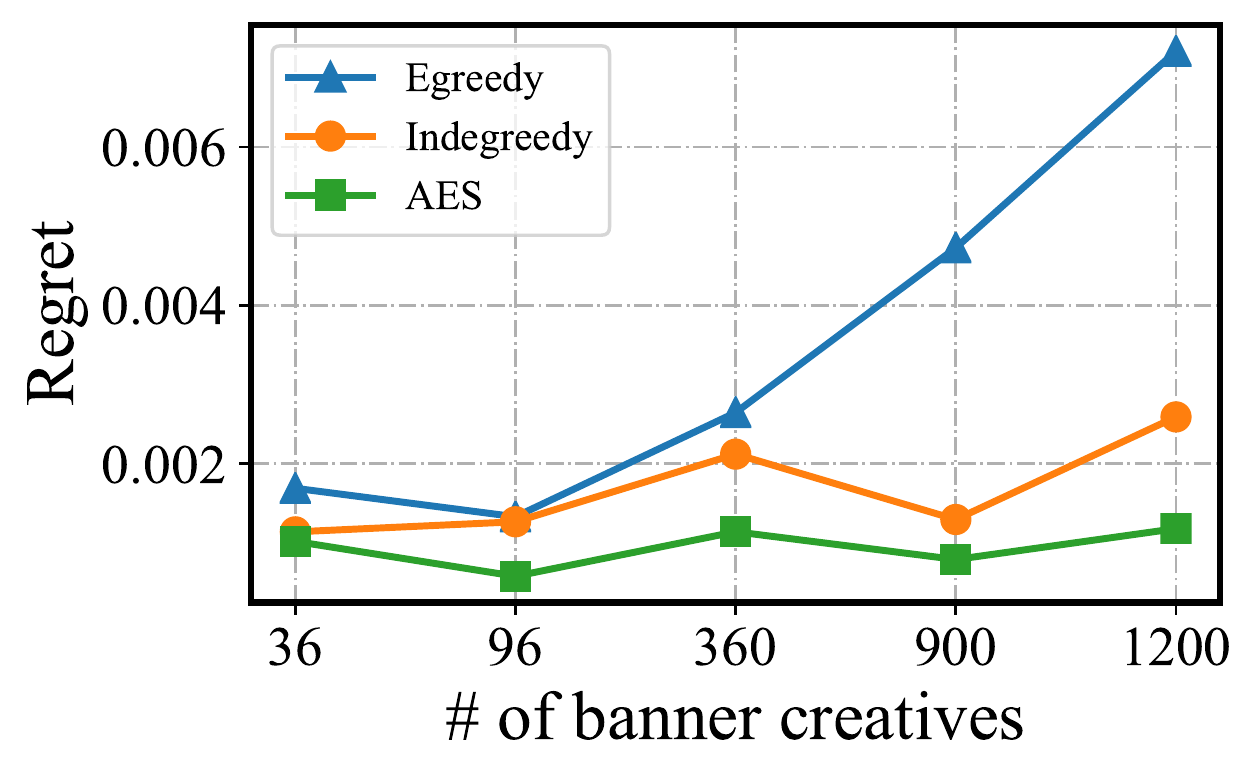}
        \caption{Algorithm performance vs. sizes of search space. Lower regret is better. 20 repetitions are run.}
        \label{regrets_diff_space}
      \end{minipage}
    \begin{minipage}{0.33\textwidth}
      \centering
      \captionsetup{margin=0.07in}
      \includegraphics[scale=0.47]{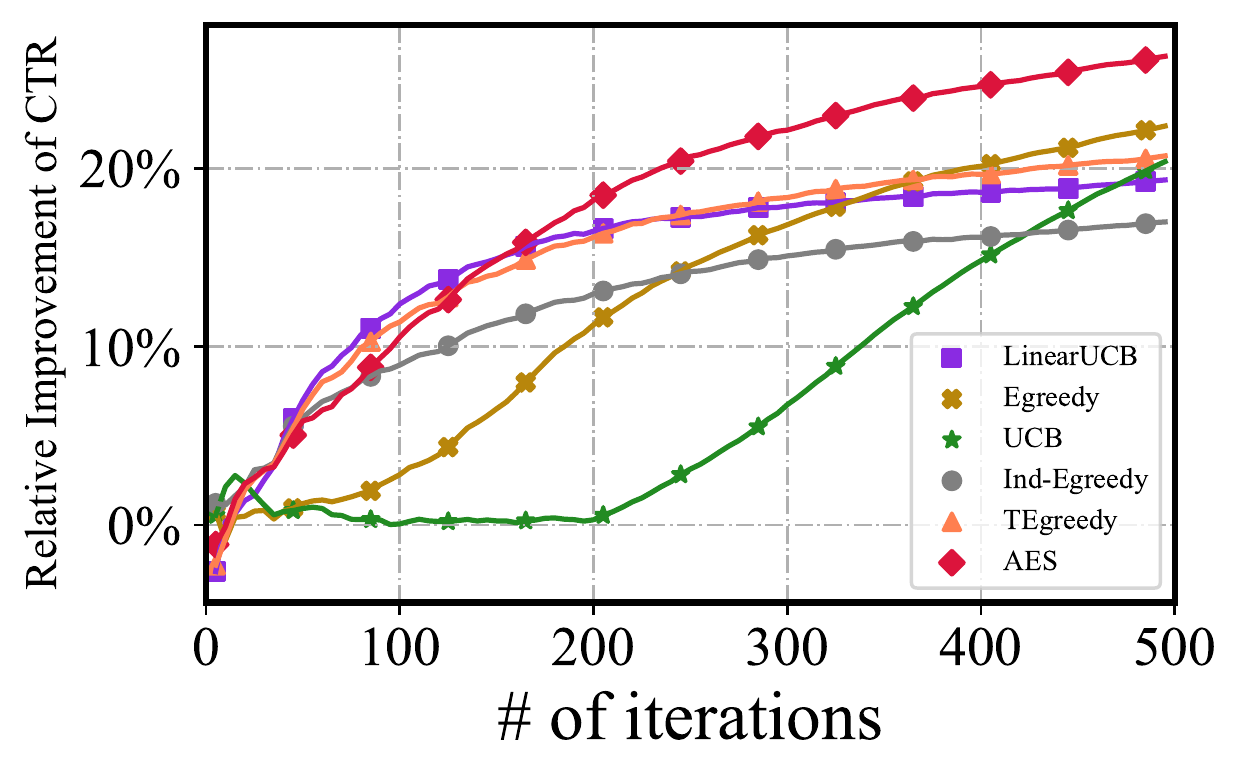}
      \caption{Experiments are run on the real-world dataset for 50 repetitions compared with \textbf{Random} policy.} 
      \label{res_online_data}
    \end{minipage}
    \begin{minipage}{0.33\textwidth}
      \centering
      \captionsetup{margin=0.07in}
      \includegraphics[scale=0.47]{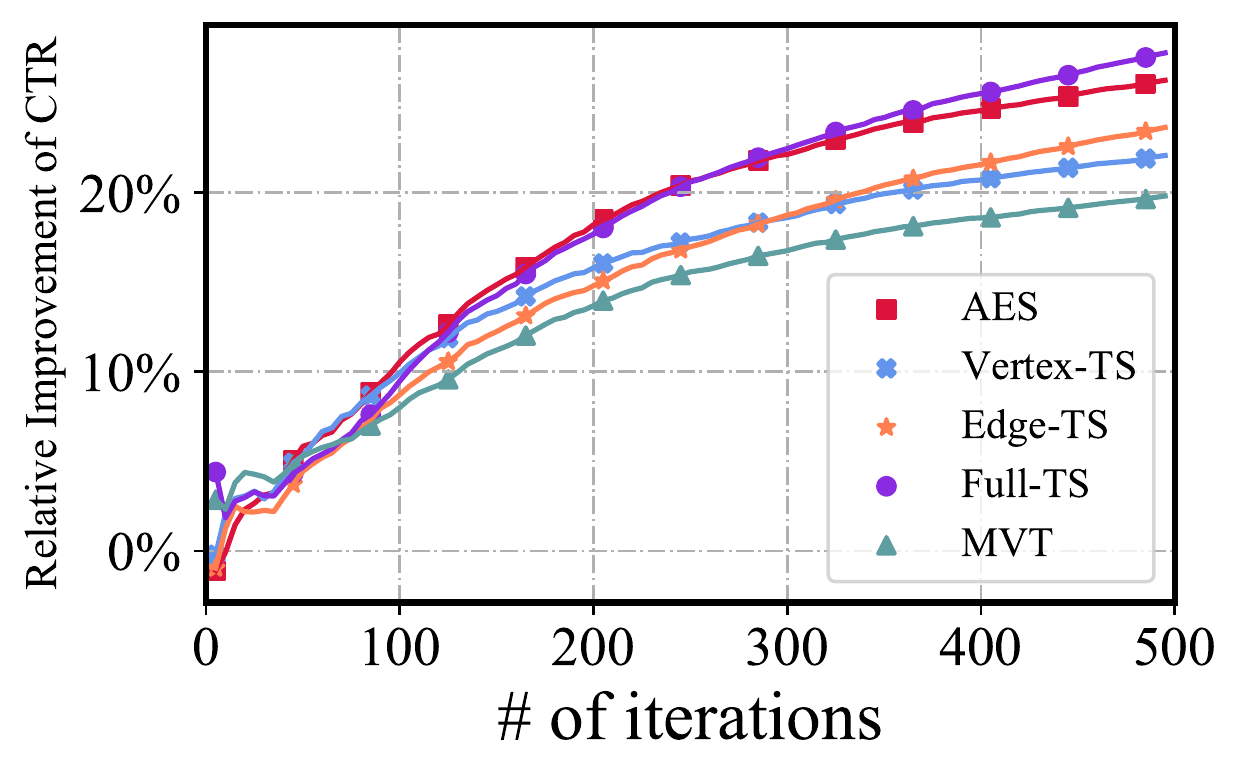}
      \caption{Experiments with different relationships of elements on the real-world dataset. Best viewed in color.}
      \label{res_different_edge}
    \end{minipage}
  \end{figure*}

\subsubsection{Sensitive Analysis for Search Space.}
We conduct experiments under different numbers of ad creatives and report the \textit{Regret} for comparisons in Fig.\ref{regrets_diff_space}. We maintain the structure of the element graph while changing the number of elements for each ingredient. The number of creatives varies from 32 to 1,200. Experiments are run in 500 batches with 30 rounds.
  
\textit{Finding4}: Our proposed method has competitive performance with lower regrets even in much larger search spaces. The regret of Egreedy dramatically increases with the growing number of candidate creatives which takes more chances to evaluate the creatives with similar expected rewards. Compared with Ind-Egreedy, our proposed method relying on both the elements and the interactions between them continuously shows superior performances under all search spaces. The tree structure increases the exploration efficiency under huge search space where relationships between elements are useful to find the optimal creative.

In conclusion, empirical experiments on synthetic data reveal that our proposed method with tree-based structures is competitive compared with regular bandit methods. The tree structure for ad creatives both accelerates the convergence and has the ability to find the optimal solution. Under much larger search spaces, our methods are still competitive with lower regrets.
  
\subsection{Experiments on Real-World Dataset}
\subsubsection{Data Collection}
We collect the online record of seven consecutive days as an offline dataset, from a relatively stable advertising space of a famous e-commerce company. About 850,000 impression logs are obtained, covering nearly 2,000 products, and are aggregated to the aforementioned 200 creatives. Thus, each creative has a statistical CTR, which is the probability of Bernoulli trials conducted for simulating user feedbacks. Every 1000 Bernoulli trials are collected to update the model and recommend next creative. Each algorithm is run for 50 rounds with 500 batches.

\subsubsection{Effectiveness}
We evaluate the performances with the following algorithms: Egreedy, UCB, LinearUCB, Ind-Egreedy, TEgreedy and AES. Results of relative CTR gain compared with the random policy are reported in Fig.\ref{res_online_data}.

\textit{Finding1}: The tree structure benefits efficiency and effectiveness for exploring the optimal creative. We focus on the convergence speed in terms of CTR since the budget for advertising is limited. Context-free baselines such as Egreedy and UCB, take more time for exploration at the very beginning so that they show worse performances. Compared with simple solutions, LinearUCB and Ind-Egreedy, our proposed method converges quickly to better solutions and continuously has better performance. This demonstrates that the relationships modeled by the ingredient tree are related to the user engagement and they help increase the overall CTR.

\textit{Finding2}: Both the elements and their interactions are helpful for accurate CTR estimation. To evaluate different performances with different types of interactions between elements, we conduct experiments with different estimators, as shown in Fig.\ref{res_different_edge}. Edge-TS only focuses on the influence of edges while Vertex-TS exploits the weights of vertices over the element graph. AES algorithm captures more information than Edge-TS and Vertex-TS. It has better performances than the two methods, which indicates the contribution both of the vertex and edge in the ad creatives on user clicks.

\textit{Finding3}: The accuracy of CTR estimator is almost not affected by the tree structure assumption. Notably, although any possible interaction between different elements is captured in Full-TS, AES still shows comparable performance but in significantly less time complexity, as shown in Fig.~\ref{res_different_edge}. This implies that the interactions of the tree-based structure play important roles in the estimation of CTR. Moreover, AES allows efficient search based on dynamic programming by introducing the tree structure while Full-TS needs to enumerate all the candidate creatives.

\begin{table}[t]
  \centering
  \begin{tabular}{c c c}
    \toprule[1.5pt]
    Method & Time(Second) & Optimal Selection \\
    \midrule[1.5pt]
    Full-TS & 729.70 $\pm$ 3.09 & Enumerated computation \\ 
    \midrule[1pt]
    MVT & 656.96 $\pm$ 2.45 & Hill Climbing \\ 
    \midrule[1pt]
    AES & \textbf{165.04 $\pm$ 0.07} & Dynamic Programming\\ 
    \bottomrule[1.5pt]
  \end{tabular}
  \caption{Comparisons of running time for searching the optimal creative on the real-world dataset. Experiments are run for 10 repetitions under 50,000 impressions.}
  \label{run_time}
\end{table}

\textit{Finding4}: The tree structure allows efficient searching for optimal creative and shows superior performances on overall CTR. As shown in Fig.\ref{res_different_edge}, compared with MVT method, which captures more interactions between elements but utilizes hill climbing for searching, our proposed method shows superior performances on CTR. The gap between Full-TS and MVT implies the inaccuracy of hill climbing for optimal selection. The ingredient tree enables accurate optimal selection under a huge number of composited creatives through dynamic programming.
Furthermore, these methods differ in the efficiency of searching optimal ad creative, as shown in Table.\ref{run_time}. We set the number of iterations $S=4$ and $K=3$ for hill climbing as mentioned in ~\cite{hill2017efficient}. AES takes the minimum time compared with hill climbing and enumerated method. That is to say, the construction of the ingredient tree significantly reduces the time complexity under huge creative space.

To sum up, the proposed method shows superior performances on the real-world dataset over competing baselines in terms of overall CTR. The accuracy of CTR estimator is not affected by the tree structure assumption and efficient searching can be implemented for optimal selection via dynamic programming.

\section{Conclusions}
This is the first work to investigate the optimal ad creative selection and we propose an \textbf{A}daptive and \textbf{E}fficient ad creative \textbf{S}election (AES) framework based on ingredient tree. The tree-based structure allows dynamic programming for efficient search under huge creative space. Due to the extremely sparse feedbacks, the exploration method, Thompson Sampling, is suggested to find potential good creatives and increase the overall CTR. To show the effectiveness of our methods, we conduct experiments both on synthetic data and real-world data and our approach shows superior performance over competing baselines.

\section*{Acknowledgements}
The work was supported by Alibaba Group through Alibaba Innovative Research Program. Defu Lian is supported by grants from the National Natural Science Foundation of China (Grant No. 61976198, 62022077),  the National Key R\&D Program of China under Grant No. 2020AAA0103800 and the Fundamental Research Funds for the Central Universities. Kai Zheng is supported by NSFC (No. 61972069, 61836007, 61832017) and Sichuan Science and Technology Program under Grant 2020JDTD0007.

\bibliography{myreference}
\end{document}